\documentclass[sigconf, nonacm]{acmart}

\usepackage{amsthm}
\usepackage{comment}
\usepackage{graphicx}
\usepackage{makecell}
\usepackage{lipsum}
\usepackage{listings}
\usepackage{subfig}

\newtheorem{theorem}{Theorem}
\newtheorem{definition}{Definition}

\newcommand\vldbdoi{XX.XX/XXX.XX}
\newcommand\vldbpages{XXX-XXX}
\newcommand\vldbvolume{14}
\newcommand\vldbissue{1}
\newcommand\vldbyear{2020}
\newcommand\vldbauthors{\authors}
\newcommand\vldbtitle{\shorttitle} 
\newcommand\vldbavailabilityurl{URL_TO_YOUR_ARTIFACTS}
\newcommand\vldbpagestyle{plain} 

\usepackage{color}

\begin{document}
\title{MatrixGate: A High-performance Data Ingestion Tool for Time-series Databases}

\author{Shuhui Wang}
\affiliation{%
  \institution{Tsinghua University}
}
\email{wangsh22@mails.tsinghua.edu.cn}

\author{Zihan Sun}
\affiliation{%
 \institution{Tsinghua University}
}
\email{sunzh22@mails.tsinghua.edu.cn}

\author{Chaochen Hu}
\affiliation{%
  \institution{Tsinghua University}
}
\email{hucc20@mails.tsinghua.edu.cn}

\author{Chao Li}
\affiliation{%
 \institution{Tsinghua University}
}
\email{li-chao@tsinghua.edu.cn}

\author{Yong Zhang}
\affiliation{%
 \institution{Tsinghua University}
}
\email{zhangyong05@tsinghua.edu.cn}

\author{Yandong Yao}
\affiliation{%
 \institution{yMatrix Inc.}
}
\email{yandong@ymatrix.cn}

\author{Hao Wang}
\affiliation{%
 \institution{yMatrix Inc.}
}
\email{magi345@163.com}

\author{Chunxiao Xing}
\affiliation{%
 \institution{Tsinghua University}
}
\email{xingcx@tsinghua.edu.cn}

\begin{abstract}
Recent years have seen massive time-series data generated in many areas. 
This different scenario brings new challenges, particularly in terms of data ingestion, where existing technologies struggle to handle such massive time-series data, leading to low loading speed and poor timeliness.

To address these challenges, this paper presents MatrixGate, a new and efficient data ingestion approach for massive time-series data. 
MatrixGate implements both single-instance and multi-instance parallel procedures, which is based on its unique ingestion strategies.
First, MatrixGate uses policies to automatically tune the slots that are synchronized with segments to ingest data, which eliminates the cost of starting transactions and enhance the efficiency. 
Second, multi-coroutines are responsible for transfer data, which can increase the degree of parallelism significantly.
Third, lock-free queues are used to enable direct data transfer without the need for disk storage or lodging in the master instance. 
Experiment results on multiple datasets show that MatrixGate outperforms state-of-the-art methods by 3 to 100 times in loading speed, and cuts down about 80\% query latency. 
Furthermore, MatrixGate scales out efficiently under distributed architecture, achieving scalability of 86\%.
\end{abstract}

\maketitle

\pagestyle{\vldbpagestyle}
\begingroup\small\noindent\raggedright\textbf{PVLDB Reference Format:}\\
\vldbauthors. \vldbtitle. PVLDB, \vldbvolume(\vldbissue): \vldbpages, \vldbyear.\\
\href{https://doi.org/\vldbdoi}{doi:\vldbdoi}
\endgroup
\begingroup
\renewcommand\thefootnote{}\footnote{\noindent
This work is licensed under the Creative Commons BY-NC-ND 4.0 International License. Visit \url{https://creativecommons.org/licenses/by-nc-nd/4.0/} to view a copy of this license. For any use beyond those covered by this license, obtain permission by emailing \href{mailto:info@vldb.org}{info@vldb.org}. Copyright is held by the owner/author(s). Publication rights licensed to the VLDB Endowment. \\
\raggedright Proceedings of the VLDB Endowment, Vol. \vldbvolume, No. \vldbissue\ %
ISSN 2150-8097. \\
\href{https://doi.org/\vldbdoi}{doi:\vldbdoi} \\
}\addtocounter{footnote}{-1}\endgroup

\ifdefempty{\vldbavailabilityurl}{}{
\vspace{.3cm}
\begingroup\small\noindent\raggedright\textbf{PVLDB Artifact Availability:}\\
The source code, data, and/or other artifacts have been made available at \url{\vldbavailabilityurl}.
\endgroup
}

\renewcommand{\thefootnote}{\fnsymbol{footnote}}

\section{Introduction}
\label{sec:introduction}
In recent years, with the rapid development of communication technology and the semiconductor industry, smart network devices are being used more and more extensively. 
These devices generate massive amounts of time-series data continuously at an unprecedented rate in various domains such as large equipment monitoring, smart cars, smart cities, environmental monitoring, telecommunications and finance. 
For example, a large Internet of Things (IoT) can consist of millions of devices, resulting in millions of records during a sample interval, which can be as low as 0.1 second\footnote{\url{https://www.electronicdesign.com/technologies/power/article/21802213/designing-ultra-low-power-sensor-nodes-for-iot-applications}}.
Over the course of an hour, billions of records would be generated.

For processing big data, a typical database mainly provides four types of functions: data loading or ingestion\footnote{In fact, both data loading and data ingestion refer to inputting data into a database and supporting manipulations or queries of these data afterwords, whereas \textit{loading} is from the data generator's perspective and \textit{ingestion} is from the database administrator's perspective. In this paper we use \textit{ingestion}.}, data storage, data query and data analysis. 
Given the rapidly growing data scenario, there is an urgent demand for key techniques in time-series databases, including: efficient data ingestion~\cite{DBLP:journals/pvldb/JensenPT18,DBLP:journals/acta/ONeilCGO96}, highly compressed data storage~\cite{DBLP:conf/usenix/VisheratinSYMNB20,DBLP:journals/imwut/BlalockMG18,DBLP:journals/pvldb/JensenPT18,DBLP:conf/sosp/AgrawalV17,DBLP:conf/icde/YuP0WSMX20}, scalable distributed architecture~\cite{DBLP:journals/pvldb/Garcia-Arellano20,clickhousecluster,DBLP:journals/tkde/JensenPT17}, near real-time querying capacity, and support for relations and transactions~\cite{DBLP:journals/pvldb/Garcia-Arellano20,DBLP:journals/pvldb/0018HQ00MFZ0ZKJ20}. 
Among these techniques, data ingestion is the most fundamental and challenging one, as the speed of ingesting data into a database must catch up with that of data generation; otherwise, more and more data would be blocked out of the database and even discarded. 
Therefore, the capability of data ingestion is the prerequisite for effective time-series data storage, query and analysis.

\begin{table*}[!htb]
\centering
\caption{Different data ingestion approaches.}
\label{tab:ingestion}
\begin{tabular}{rccc}
\hline
Approach & SISP & SIPP & MIPP \\ \hline
Instance & Single & Single & Multiple \\
Degree Of Parallelism (DOP) & 1 & Limited & High \\
Throughput & Low & Medium & High \\
Scalability & Poor & Poor & Good \\
CPU utilization & Low & High in one host & High in the cluster \\
Timeliness & Good & Poor & Poor \\
Examples & \makecell{INSERT, COPY or\\LOAD~\cite{momjian2001postgresql,dubois2008mysql,mysql/insert,mysql/load,postgresql/copy}} & \makecell{INSERT in ClickHouse~\cite{clickhouse/insert},\\\textit{timescaledb-parallel-copy}~\cite{timescaledb/parallel/copy}} & \makecell{COPY ON SEGMENT~\cite{copy/on/segment},\\external tables~\cite{greenplumExternaltable,oracleExternaltable}} \\
 \hline
\end{tabular}
\end{table*}

Time-series data has three main characteristics~\cite{6949290,nerc527832}: 
1. Each record is timestamped and can be uniquely identified by its timestamp and device ID;
2. The measurement value is fixed after generated at a certain point of time, meaning time-series data is not to be updated;
3. The amount of data grows very fast as the data generating devices produce large amounts of data at a certain sampling frequency continuously.
Since traditional relational databases require stronger data consistency, they are not suitable for processing such data. 
As a result, over the past decades, time-series databases have developed rapidly and in the last two years it has become the most active and fastest growing subfield in the database industry due to the continuous, fast and massive growth of time-series data and the urgent needs in various application areas~\cite{dbengine}.

\begin{figure}[!htb]
    \centering
        \subfloat[SISP]
        {
            \centering
            \includegraphics[width=.4\textwidth]{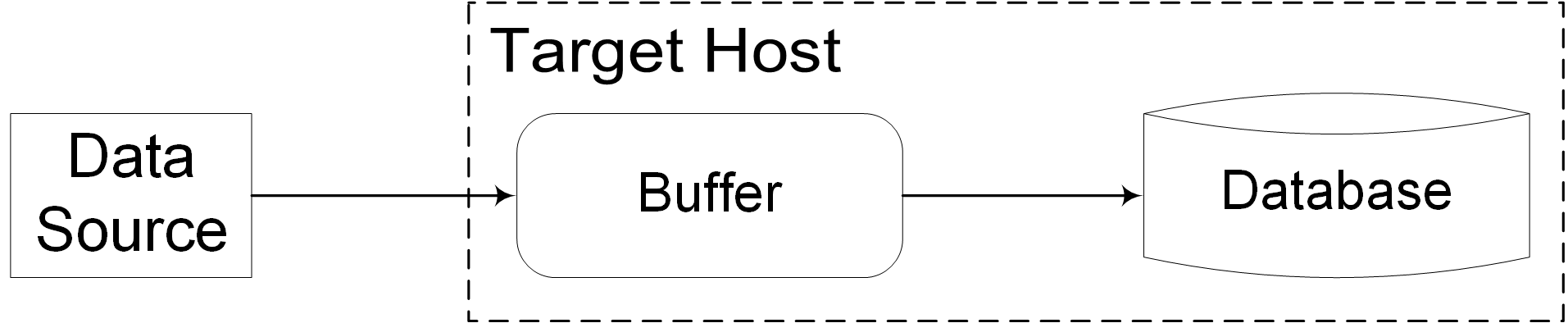}
            \label{subfig:sisp}
        }
        
        \subfloat[SIPP]
        {
            \centering
            \includegraphics[width=.4\textwidth]{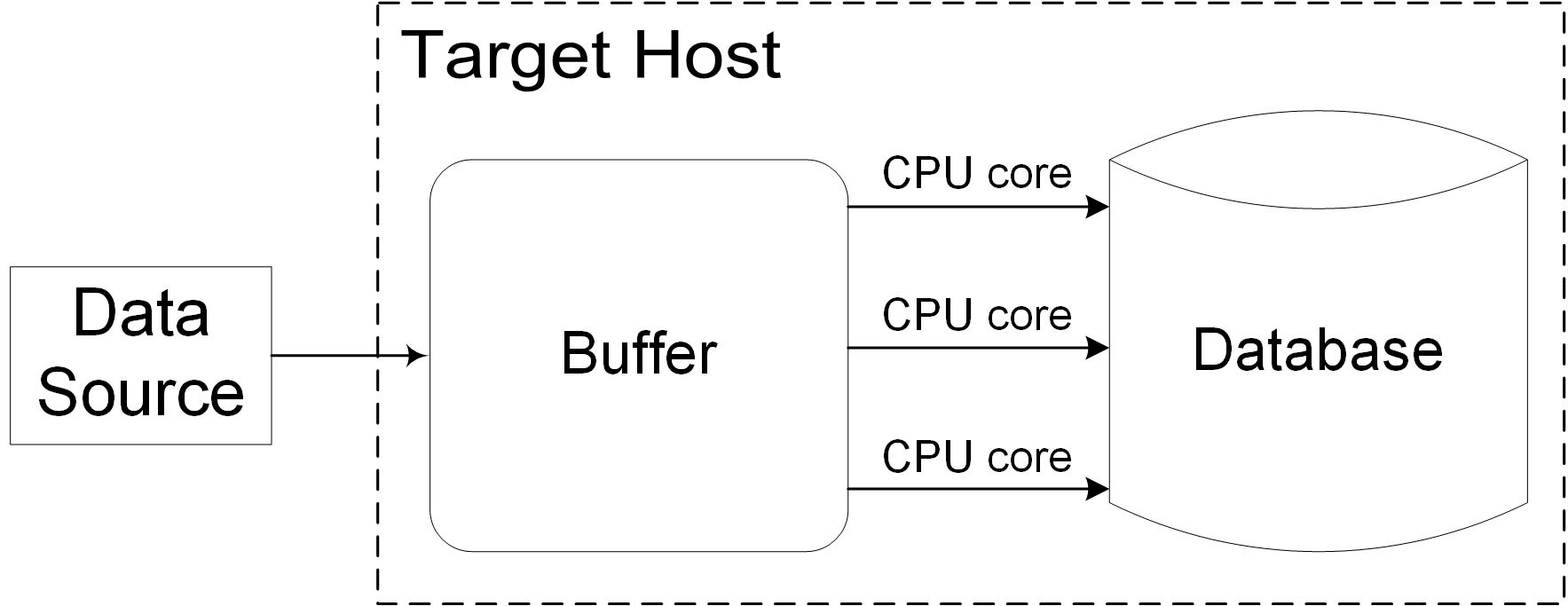}
            \label{subfig:sipp}
        }
        
        \subfloat[MIPP]
        {
            \centering
            \includegraphics[width=.4\textwidth]{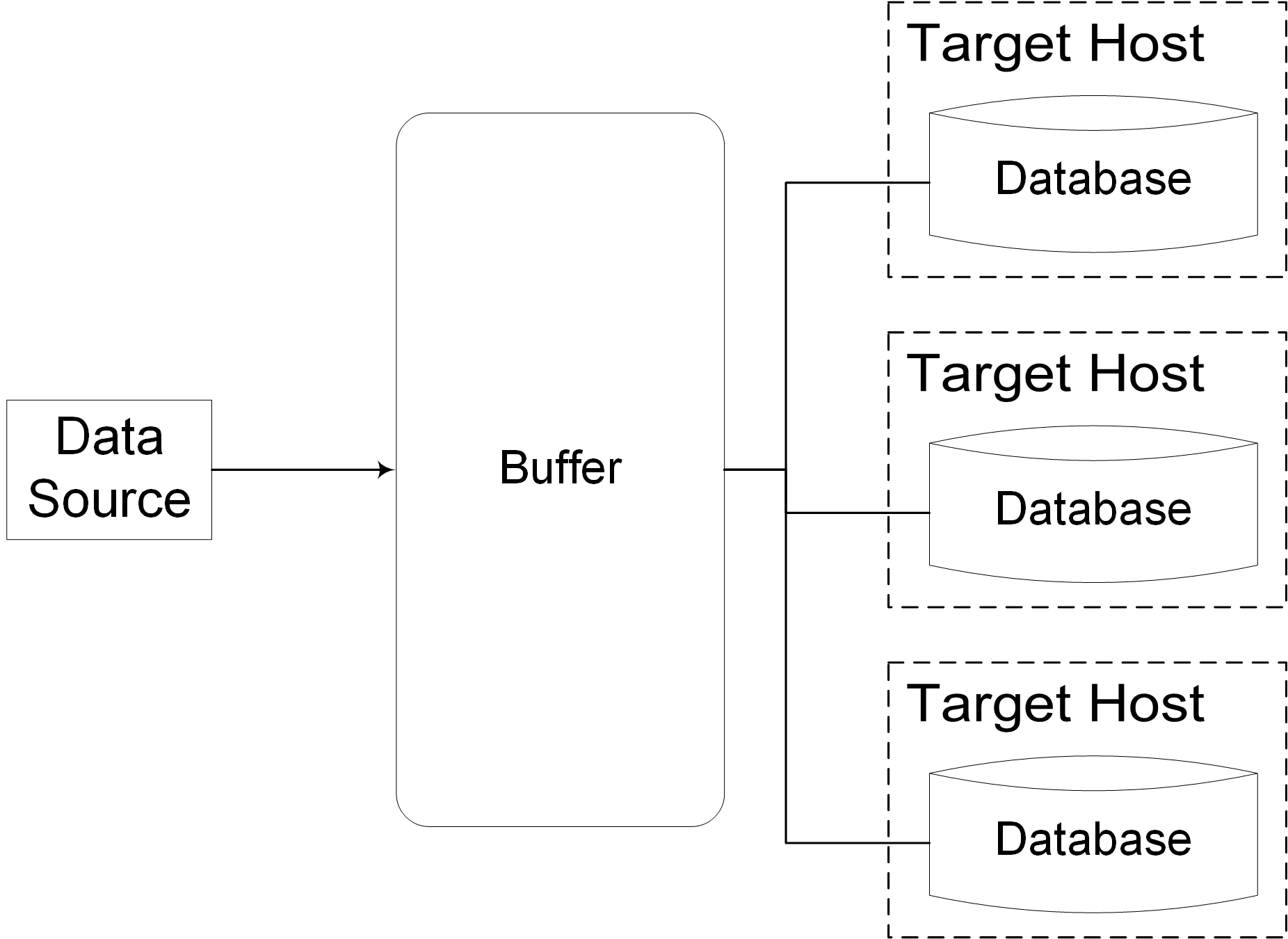}
            \label{subfig:mipp}
        }
    \caption{Different ingestion approaches.}
    \label{fig:ingestion}
\end{figure}

Existing data ingestion approaches can be divided into three categories, \textit{Single-instance serial-procedure} (SISP), \textit{Single-instance parallel-procedure} (SIPP) and \textit{Multiple-instance parallel-procedure} (MIPP). 
SISP, or the naive approach, is just to ingest data serially, which is extremely low in both throughput and efficiency, thus rarely used to load large-scale data. 
The other two approaches will be discussed in Section~\ref{sec:relatedWork}. 
In general, Figure~\ref{fig:ingestion} shows the process of three approaches, and Table~\ref{tab:ingestion} presents their performance.

Table~\ref{tab:ingestion} depicts that existing time-series data ingestion approaches are still inadequate in terms of throughput and timeliness when meeting the challenges posed by the rapidly growing amount of time-series data~\cite{reinsel2017data}. 
In addition, all of the approaches use a buffer to stage received data, introducing additional overhead if data needs to be written to disk.

Our proposed solution, MatrixGate, leverages parallel process to achieve efficient parallel ingestion on a single instance and supports horizontal scaling on distributed architectures.
That is to say, MatrixGate can achieve higher throughput on multiple hosts. 
In order to solve the issues of low loading speed and poor timeliness, MatrixGate commits data via automatically tuning slots, on which we design a new ingestion strategy for it. 
Different from traditional multi-process and multi-thread approaches, MatrixGate implements parallel procedure via multi-coroutine, cutting down extra overhead on scheduling processes or threads since scheduling coroutines is much faster. 
Additionally, MatrixGate builds data pipelines based on lock-free communication, eliminating the need for intermediate data staging or dumping on disk. 
These data pipelines connect directly to sub-instance, bypassing the main database instance, avoiding the single point bottleneck problem. 
These designs ensure the outperformance of MatrixGate.

In summary, we make the following contributions.

\begin{enumerate}
\item We build MatrixGate, a high-performance time-series data ingestion tool.
\item We provide automatically tuning slots that are synchronized with segments to ingest data, which eliminates the cost of starting transactions and enhance the efficiency.
\item We employ multi-coroutine instead of multi-process or multi-thread to implement parallelism, reducing overheads on scheduling processes or threads.
\item We achieve direct data transfer by using lock-free communication, so that no data need to be written to disk during data transfer.
\item We conduct extensive experiments on four typical time-series datasets, and the experiments results validate the high performance of MatrixGate.
\end{enumerate}

\begin{figure*}[!htb]
    \centering
    \includegraphics[width=\textwidth]{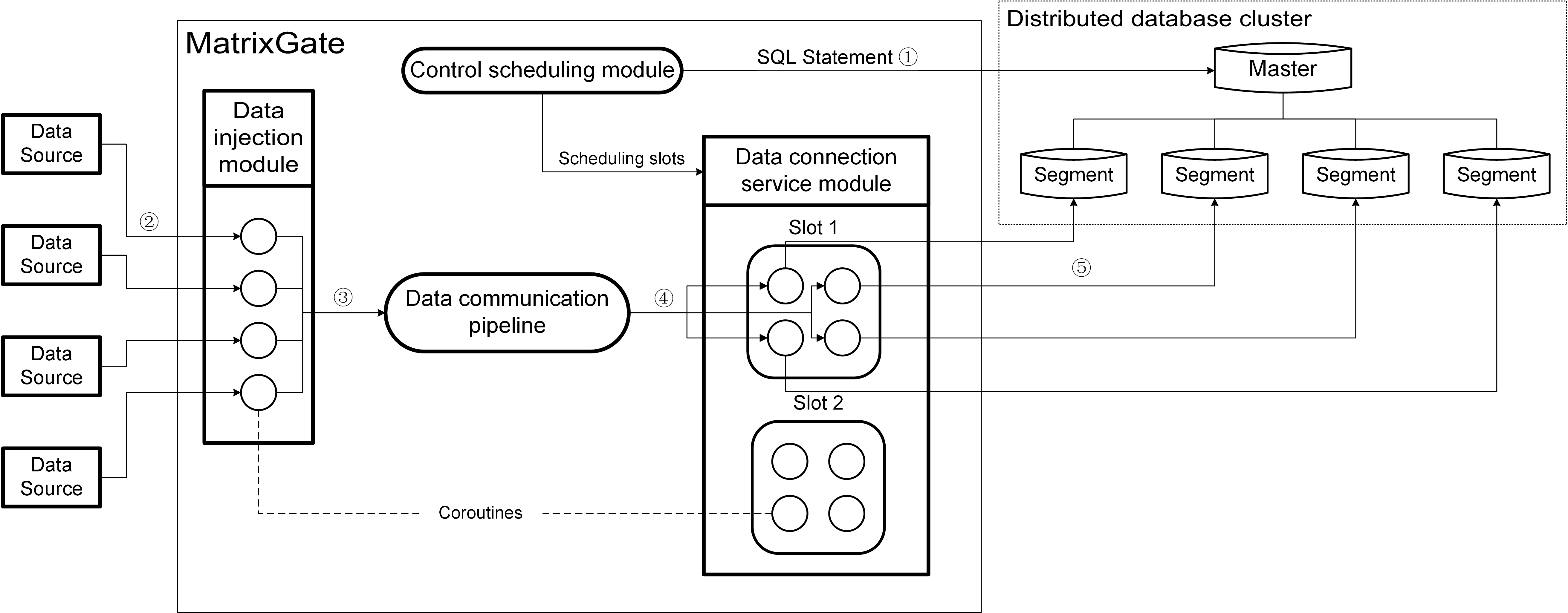}
    \caption{The architecture of MatrixGate.}
    \label{fig:GateArch}
\end{figure*}

The structure of this paper is as follows: 
Section~\ref{sec:matrixGate} describes MatrixGate's architecture and components. 
Then, Section~\ref{sec:design} and Section~\ref{sec:detail} present its features. 
Section~\ref{sec:experiment} reports our evaluation results, while Section~\ref{sec:relatedWork} discusses about existing approaches. 
Finally, Section~\ref{sec:conclusion} concludes the paper.

\section{Architecture}
\label{sec:matrixGate}
The general architecture of MatrixGate is shown in Figure ~\ref{fig:GateArch}. 
The components of MatrixGate are the following four modules. 
The \textbf{control scheduling module} schedules other modules of MatrixGate, and is responsible for connecting to and controlling the target database so that each module can coordinate to load time-series data. 
The \textbf{data injection module} is responsible to listen at a certain IP port, where data collection devices generate data. 
The \textbf{data communication pipeline}, based on lock-free queues~\cite{valois1994implementing,DBLP:conf/podc/MichaelS96,DBLP:conf/wdag/Ladan-MozesS04}, allows multiple coroutines to read and write simultaneously and connects the data injection module with the \textbf{data connection service module}. 
It is responsible for connections to database sub-instances, initiating coroutines to dock the database sub-instances and transferring data to the database sub-instances.

\begin{definition}
A segment refers to a sub-instance of the database.
\end{definition}
\begin{definition}
A slot is a loader instance of the \textit{data connection service module} that is responsible to transfer data to segments. At any given moment, only one slot is active.
\end{definition}

During data ingestion, MatrixGate follows the steps in Figure ~\ref{fig:GateArch}. 

\begin{enumerate}
    \item When an ingestion procedure begins, the \textit{control scheduling module} first establishes a connection to the database, initiates and maintains a transaction to insert data from an external table.
    \item The coroutines of the \textit{data injection module} listen to a specific address, from which the data collection device posts data.
    \item The coroutines of the \textit{data injection module} put the collected data into the \textit{data communication pipeline}.
    \item The coroutines of the \textit{data connection service module} load data from the \textit{data communication pipeline}.
    \item The coroutines of the \textit{data connection service module} transfer the data directly to the segments. After a certain interval, the coroutines send an EOF to the segments to end the connection, before another batch of coroutines start the next ingestion procedure.
\end{enumerate}

As we mentioned in Section~\ref{sec:introduction}, the parallel operation of MatrixGate is implemented via coroutines. 
Specifically, the \textit{data injection module} and \textit{data connection service module} are all made up of numerous coroutines.
The following description of MatrixGate is divided into two sections.
Section~\ref{sec:design} introduces the design of ingestion process and Section~\ref{sec:detail} introduces the implementation details of MatrixGate.

\section{Method}
\label{sec:design}
We will discuss the architecture further in this section. 
The section consists of two parts: 
Section~\ref{subsec:strategy} describes the new strategy used by MatrixGate compared to traditional databases, and section~\ref{subsec:slot} describes the details concerning about scheduling slots in our design.

\begin{figure*}[!htb]
    \centering
    \includegraphics[width=.7\textwidth]{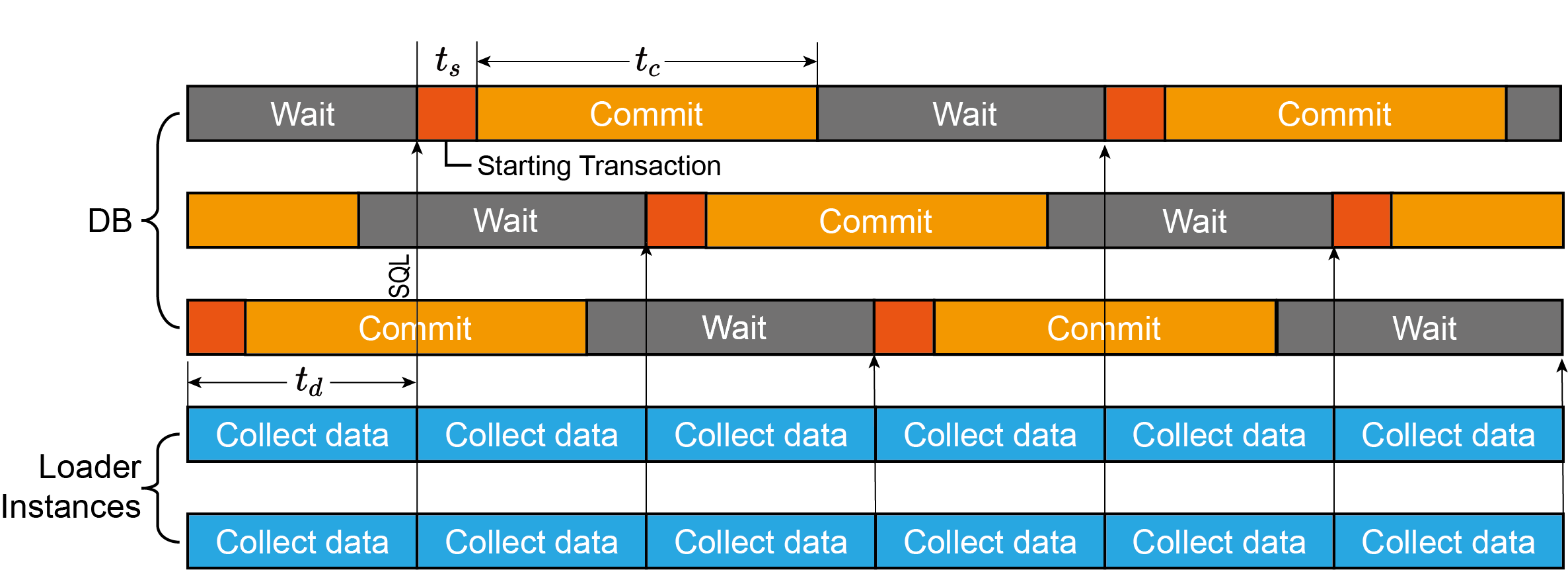}
    \caption{The timing sequence of naive data ingestion.}
    \label{subfig:ganttTra}
\end{figure*}

\begin{figure*}[!htb]
    \centering
    \includegraphics[width=\textwidth]{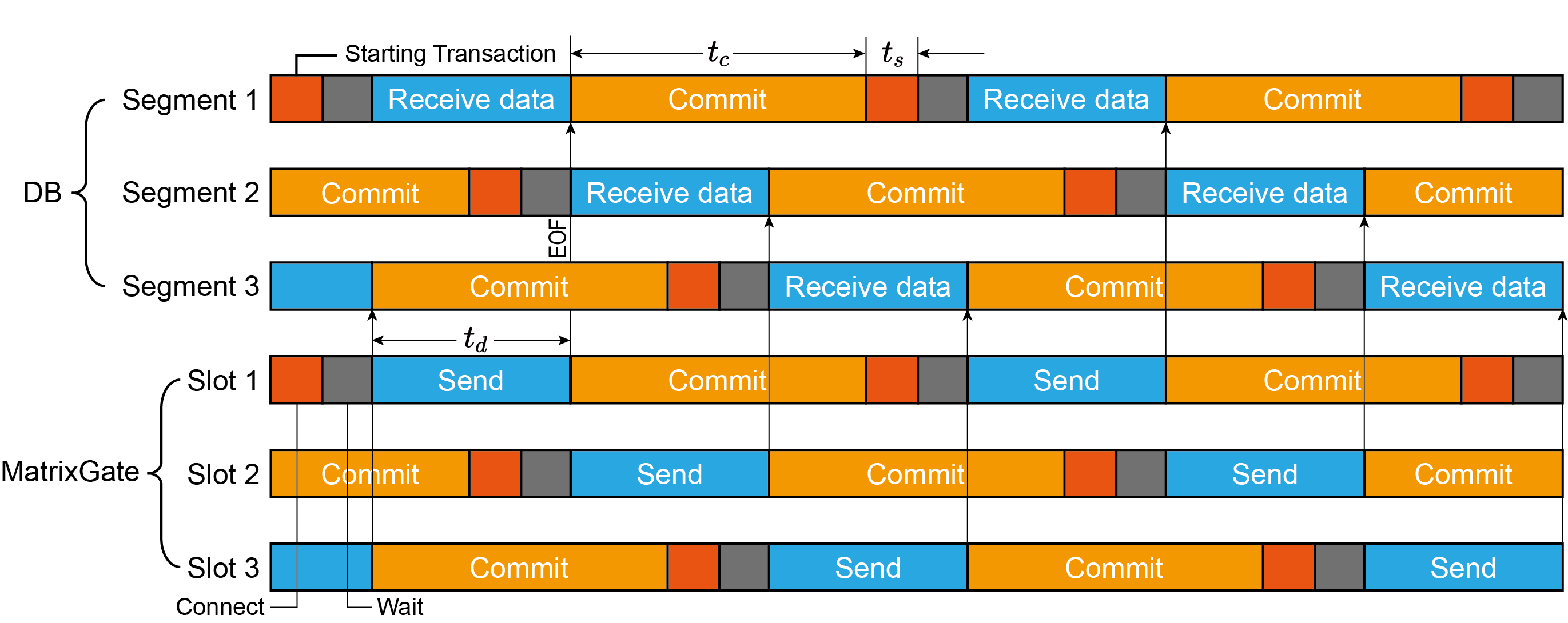}
    \caption{The timing sequence of MatrixGate's data ingestion.}
    \label{subfig:ganttMat}
\end{figure*}

\subsection{Strategy}
\label{subsec:strategy}
The naive procedure of data ingestion is simple, as shown in Figure~\ref{subfig:ganttTra}. 
First loader instances collect data, probably in parallel. 
The time-series data is generated continuously, so the loader instance cannot commit to the database once it receives data; otherwise, it will commit so many times that the database cannot handle.
Then the loader instance initiates a SQL statement, and the database starts a corresponding transaction. 
Hence, the time span from when the data arrives at the loader instance to when the data is visible to users can be calculated as:
\[T=t_d+t_s+t_c\]
where $t_d$ is the time of collecting data, $t_s$ is the time of starting the transaction, and $t_c$ is the time of committing the transaction. 

The key idea of our MatrixGate is that $t_s$ can be eliminated during the ingestion process by starting the transaction in advance, as shown in Figure~\ref{subfig:ganttMat}. 
As mentioned in Section~\ref{sec:matrixGate}, the \textit{data connection service module} is responsible for ingesting data into the database. 
Each slot contains multiple coroutines and connects to one or multiple database segments, while the slot only connects to one segment in figure~\ref{subfig:ganttMat} for simplification.
In order to know the status of the segments, the slots should be synchronized with the segments. 
In MatrixGate, a slot cycle has four phases, each phase corresponding a phase of a segment. 
\begin{enumerate}
    \item \textbf{Connect}. The slot connects to its segments.
    \item \textbf{Wait}. The slot waits until the \textit{control scheduling module} orders it.
    \item \textbf{Send}. The slot sends data to its segments.
    \item \textbf{Commit}. The slot waits until the database finishes the transaction.
\end{enumerate}
During the ingestion procedure, only one slot is sending data to its segments.

After activated by the \textit{control scheduling module}, a slot begins in the connect phase and executes an SQL statement to start the transaction through the \textit{control scheduling module}. 
The data transfer is implemented through HTTPS~\cite{rescorla2018transport}, which ensures the data can be transported to the database reliably. 
MatrixGate can also detect data in wrong format and record error logs, which will not influence the ingestion of other data. 
An example of the SQL statement is given below,
\begin{lstlisting}[language=SQL]
    INSERT INTO table1 SELECT * FROM ext_table
\end{lstlisting}
where the \texttt{ext\_table} is the name of a virtual external table, which is pointed to the slot.
 
After the segments are ready to receive data, the \textit{control scheduling module} marks the slot in the wait phase, and the slot begins waiting for instructions from the module. 
When all of the other slots are neither sending data nor ready to send data, the \textit{control scheduling module} lets the slot start to send data and enter the send phase. 
The time that a slot can be used to send data, or $t_d$, is called \textbf{interval} in MatrixGate. 
The data collected by the slot during the interval is called \textbf{micro-batch}. 
After the interval ends, the slot sends an EOF to the segment and enters commit phase. 
The slot will keeps its phase until the database commits the transaction and frees the segments, and will enter the connect phase again. 
During the cycle, only the phase change from the send phase to the commit phase is executed by the slot, and the other changes are all marked by the \textit{control scheduling module}, which also communicates with the segments. 
In Figure~\ref{subfig:ganttMat}, the slot connects its old segments again, while in fact the slot can connect to other segments. 

In conclusion, by synchronizing the status between slots and segments, the \textit{control scheduling module} can check the status of segments and know which segments are free to start transactions. 
Hence, the preparation of the transaction can be done before data transfer. 
The time span from when the data arrives at the loader instance to when the data is visible to users can be calculated as:
\[T=t_d+t_c\]
in which we eliminate the $t_s$ term. 
Under normal scenario, the $t_s$ term is minimal and can be ignored. 
However, given our scenario where the data is generated very fast, $t_d$ is restricted to some milliseconds, and also the transaction is very simple (only an INSERT statement), so $t_c$ is also small. 
In this case, $t_s$ becomes more significant and it is valuable to optimize it.

\subsection{Scheduling slots}
\label{subsec:slot}
Under our time-series scenario, data generation can maintain a variable speed, that is,
the data may not be generated at a certain speed. 
In this case, $t_d$ is preset and fixed, $t_s$ does not vary so largely since the database is assumed to be efficient, so the determining variable is $t_c$. 
The time cost of committing a transaction depends on the number of tuples it contains, so $t_c$ is correlative to the speed of data generation positively as $t_d$ is fixed.

Given changed parameters, the number of slots cannot be fixed. 
If the number of slots is too low, the \textit{control scheduling module} may be unable to find a free slot to continue data ingestion; if the number of slots is too high, there may be many free slots connecting to segments. 
As mentioned in Section~\ref{subsec:strategy}, a slot which is in the wait phase does not mean it is free, but busy with connecting segments. 
Therefore, multiple free slots with their segments lead to extra cost. 
In ideal case, there is always one free slot (as shown in~\ref{subfig:ganttMat}) in a single coroutine of the \textit{data connection service module}.

Suppose there are $N$ slots, we can get the optimal number of slots using the following theorem.
\begin{theorem}
    \label{thm:slot}
    The optimal number of slots $N^*$ is given by:
    \[N^*=\left\lceil\frac{t_d+t_c+t_s}{t_d}\right\rceil\]
\end{theorem}

\begin{proof}
First, we prove that $N<N^*$ is not feasible. 
To assure that the ingestion process is continuous, during the commit and connect phase of a slot, there must be other slots responsible for ingesting data. 
The busy time of a slot is $t_c+t_s$, so the number of extra slots, $N-1$ must not be less than $\left\lceil\frac{t_c+t_s}{t_d}\right\rceil$ and $N$ must not be less than $N^*$.

Second, we prove that using $N^*$ slots is feasible. 
We number the slots from 1 to $N^*$. 
Suppose at $T_0$ slot 1 begins to send data, and then at $T_1=T_0+N^*t_d$, slot $N^*$ ends sending data. 
On the other hand, at $T_2=T_0+t_d+t_c+t_s$, slot 1 has entered the wait phase. 
It is easy to show that $T_1\ge T_2$, so when slot $N^*$ ends sending data, slot 1 is free and can be assigned to send data. 
Similarly, when slot $1, 2, \dots, N^*-1$ ends sending data, slot $2, 3, \dots, N^*$ can be assigned to send data respectively. 
In conclusion, using $N^*$ slots is feasible to assure persistent ingestion.
\end{proof}

As we prove the optimal number of slots, we need to keep $N$ equal to or near $N^*$. 
However, we cannot just calculate $N^*$ and adjust $N$ to $N^*$, because $t_c$ and $N^*$ can change abruptly due to change of data flow. 
Therefore, we need to propose a heuristic algorithm to adjust the number of slots.

MatrixGate is able to tune the number of slots automatically and heuristically, which is conducted by the \textit{data connection service module}. 
The tuning policies used by MatrixGate are shown below.
\begin{enumerate}
    \item There is only one slot at first.
    \item When the data comes to the \textit{data connection service module}, if there are no slots in the \textit{send} phase, and neither (3) nor (4) fits, the module will activate a new slot.
    \item After activating a slot, a new slot cannot be activated until one interval ($t_d$) later.
    \item A new slot cannot be activated until the last activated slot enters its \textit{send} phase.
    \item A slot is aborted once it is in the \textit{wait} phase for more than one interval if (7) does not fit.
    \item During a dispatch cycle (default value is 10 seconds), if a slot does not send any data and (7) does not fit, it will be aborted.
    \item If there is only one slot, the slot cannot be aborted until its \textit{send} phase ends.
\end{enumerate}

Policy (1) defines the initial state of the \textit{data connection service module}, with only one slot communicating with the database. 
Policy (2) is natural since if there is no extra slot, we cannot let the data wait for working slots.

\begin{figure}[!htb]
    \centering
    \includegraphics[width=.4\textwidth]{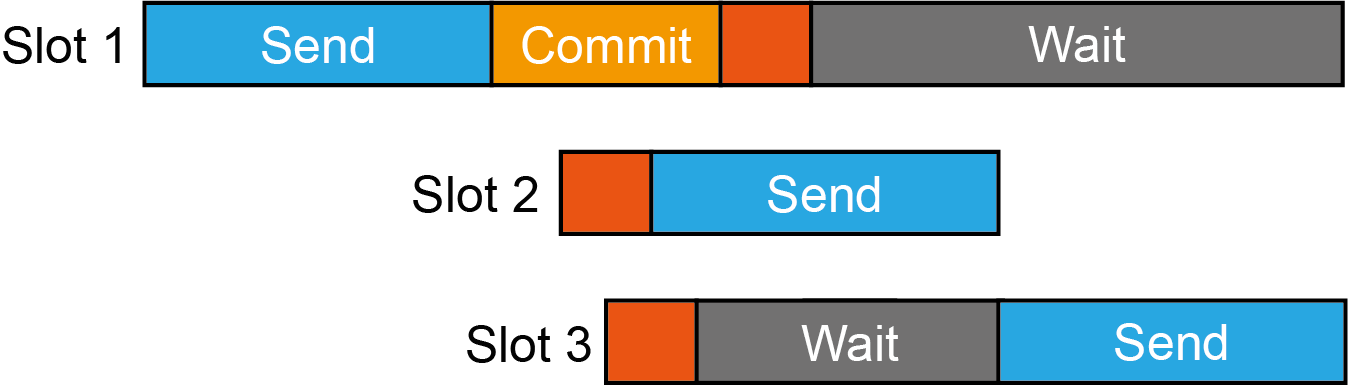}
    \caption{An example of violation of policy (3) causing unnecessary slot.}
    \label{fig:p2}
\end{figure}

Policy (3) and (4) ensures no more unnecessary slots to be generated. 
In Figure~\ref{fig:p2}, the activation of Slot 3 violates policy (3). 
As a result, although Slot 3 is ready to send data, its work is seized by Slot 2. 
After Slot 2 completes sending its micro-batch, Slot 1 has already entered the wait phase, so Slot 3 becomes unnecessary and wastes resources. 
Hence, Policy (3) and (4) ensure that only after the last activated slot ends sending data, a new slot can be activated, which can prevent generating extra slots.

Policy (5) eliminates unnecessary slots. Consider the slot cycle when $N>N^*$. 
Suppose $N=N^*+n$ where $n\ge1$. 
At $T_0$ slot 1 begins to send data. 
According to the proof of Theorem~\ref{thm:slot} above, at $T_1^\prime=T_0+Nt_d$, slot $N$ ends sending data, and at $T_2$ slot 1 enters the wait phase. 
By $T_1^\prime$, slot 1 has been in the wait phase for
\[T_1^\prime-T_2=nt_d+(T_1-T_2)\ge t_d \]
i.e., slot 1 has been in the wait phase for no less than an interval.\footnote{In this case, all the slots are in the wait phase for no less than an interval. 
If we changes dispatching method, there can be some slots that is in the wait phase for less than an interval, but there must be a slot that is in the wait phase for no less than an interval.}
That is to say, when the number of slots is too high, there exists a slot that is in the wait phase for more than an interval. 
So such a slot can be seen as a characteristic of too many slots. 
Hence, Policy (5) checks if such a slot exists and kills the slot. 
After killing the slot, time that the other slots spend on the wait phase will also decrease, so Policy (5) will not kill all the slots at one time.

Policy (6) is to deal with no data. 
With the five policies above, when there is no data, $t_c=0$ and $N^*=\left\lceil\frac{t_d+t_s}{t_d}\right\rceil=2$ because the database is free and $t_s$ is minimal. 
So two slots remain, connect and send in turn, even though there is no data to send. 
Starting and committing empty transactions definitely consumes and wastes resources. 
To prevent this, Policy (6) defines a threshold that the activity of a slot must reach. 
If there is no data during a dispatch cycle, all of the slots will be aborted gradually and the \textit{data connection service module} will stop working ultimately.

It should be noted that Policy (6) may cause to lose data.
Suppose some data enters the \textit{data connection service module} in the last moment of a dispatch cycle. 
Policy (2) shows there is one slot in the send phase, but when the data enters the slot, the slot is aborted due to Policy (6). 
In this case these data will be lost. 
Policy (7) is designed to prevent this.

\begin{figure}[!htb]
    \centering
        \subfloat[Before adding a slot.]
        {
            \centering
            \includegraphics[width=.45\textwidth]{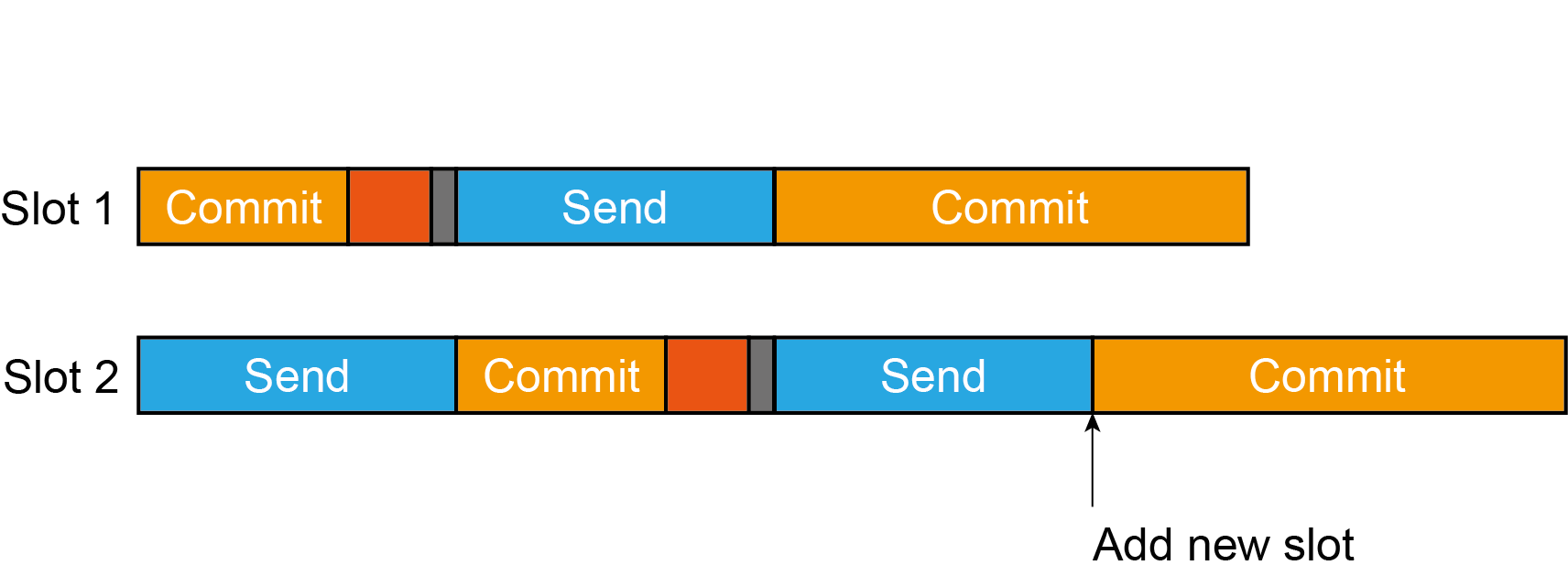}
            \label{subfig:adding}
        }

        \subfloat[After adding a slot.]
        {
            \centering
            \includegraphics[width=.45\textwidth]{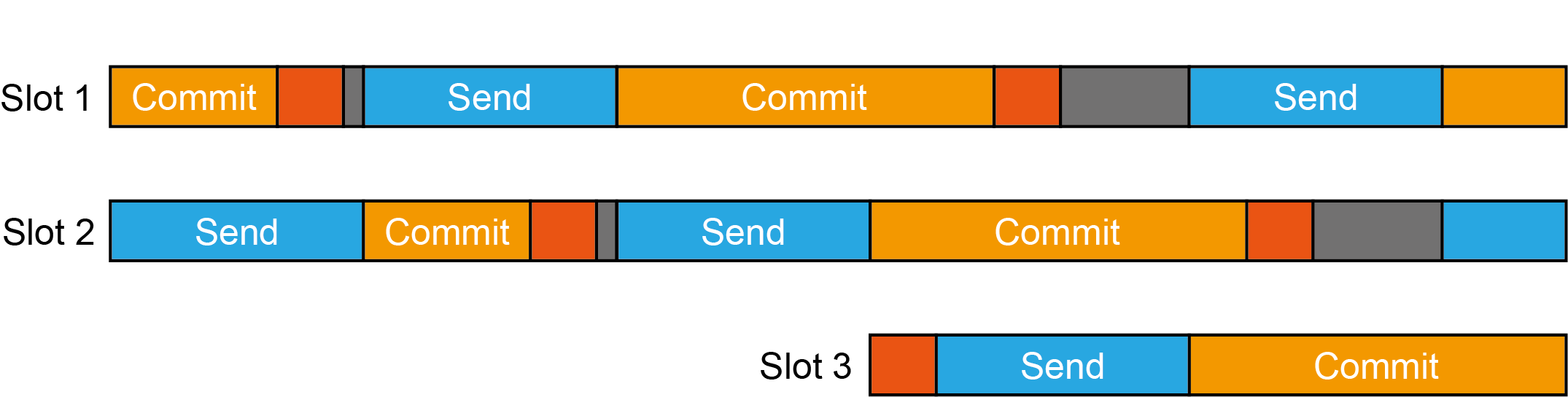}
            \label{subfig:added}
        }

        \subfloat[Before aborting a slot.]
        {
            \centering
            \includegraphics[width=.45\textwidth]{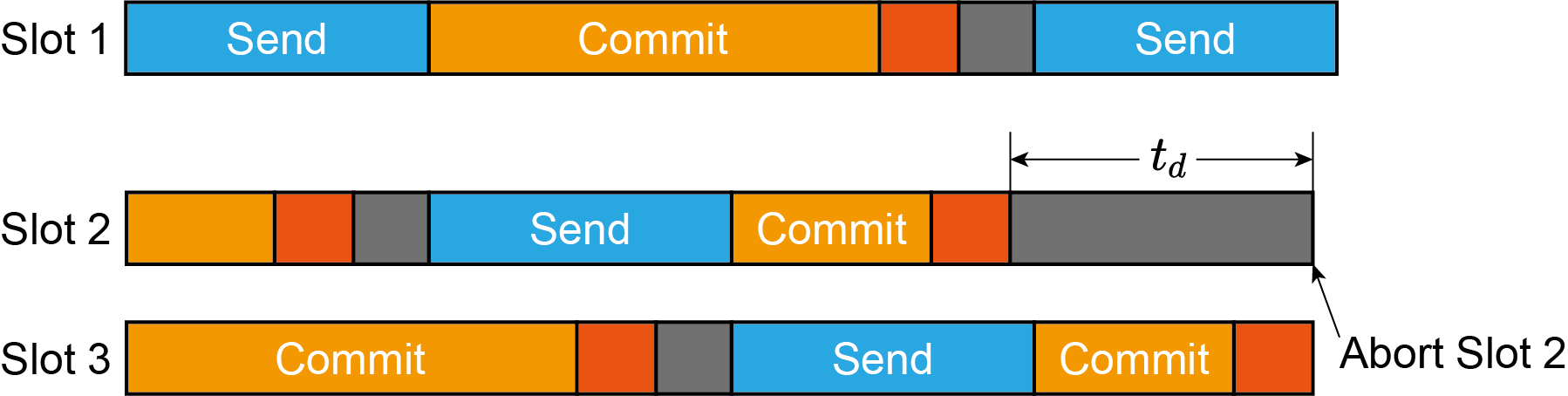}
            \label{subfig:removing}
        }

        \subfloat[After aborting a slot.]
        {
            \centering
            \includegraphics[width=.45\textwidth]{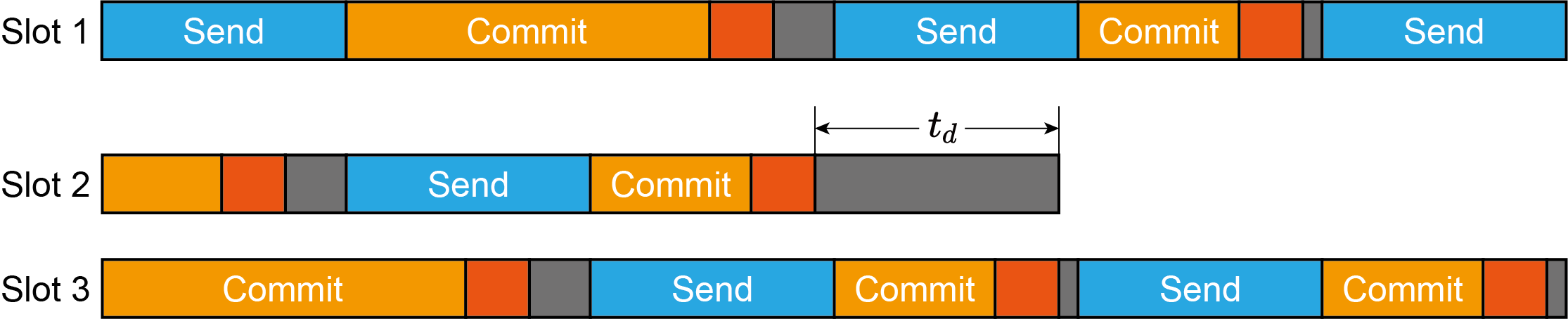}
            \label{subfig:removed}
        }
    \caption{Examples of dispatching slots.}
    \label{fig:dispatch}
\end{figure}

Now we give examples of scheduling slots, as shown in Figure~\ref{fig:dispatch}. 
Suppose there are two working slots at first. 
At some time, the data flow increases and the data collected in an interval also increases. 
As a result, $t_c$ increases, and two slots can no longer support the ingestion process. 
In Figure ~\ref{subfig:adding}, after slot 2 ends its send phase and data continues to enter the \textit{data connection service module}, the \textit{control scheduling module} finds no free slot, so it starts a new slot, namely slot 3 in Figure~\ref{subfig:added}. 
According to Policy (3) and (4), no slots will be activated until Slot 3 ends its send phase. 
As a result, data will lodge in the memory temporarily when slot 3 is connecting to its segments and transfer to the segments until it ends its connect phase. 
After the dispatching, the number of slots increases to 3, which is also the optimal number of slots.

Now we consider when data flow decreases. 
In Figure~\ref{subfig:removing}, $t_c$ drops dramatically due to decreased data flow. 
After slot 3 ends its send phase, the \textit{control scheduling module} finds two free slots: slot 1 and slot 2. 
Assume that the \textit{control scheduling module} selects slot 1 and slot 2 will not enter its send phase until slot 1 ends its send phase. 
This causes slot 2 to wait for more than an interval, and slot 2 is aborted after one interval. 
Again, the number of slots decreases to 2, which is also the optimal number of slots, as shown in Figure~\ref{subfig:removed}.

\section{Implementation Details}
\label{sec:detail}
This section will discuss two features of MatrixGate's implementation. Section~\ref{subsec:coroutine} discusses multi-coroutine and its advantage against multi-process and multi-thread. 
Section~\ref{subsec:lockfree} describes the lock-free queue used by the data communication pipeline.

\subsection{Multi-coroutine}
\label{subsec:coroutine}
Most parallel programs implement concurrency by using multiple processes or multiple threads. MatrixGate, instead, uses multiple \textbf{coroutines} to achieve high DOP. 
The parallel modules of MatrixGate are the \textit{data injection module} and the \textit{data connection service module}, so subroutines of these two modules are all coroutines. 
The concept of a coroutine is logical: a coroutine is defined as a subroutine that acts independently from other subroutines~\cite{DBLP:journals/cacm/Conway63,DBLP:books/sp/Marlin80}. 
In practice, coroutines refer to subroutines that are hidden to the operating system and whose scheduling is manually implemented by users. 
Popular programming language including Golang~\cite{golangCoroutines}, Java~\cite{stadler2011coroutines}, Lua~\cite{de2004coroutines} and Python~\cite{pythonCoroutines} all implement their own coroutines. 
MatrixGate is written by Golang, so the coroutines that MatrixGate uses are also \textit{goroutines} that Golang implements. 
We will now discuss the difference among process, thread and coroutine.

First, consider the case that DOP ($M$) is less than or equal to the number of CPU cores ($N$). 
In this ideal case, each subroutine is running in an independent CPU core in parallel, so the difference of the implementations is on the resources. 
In multi-process implementation, $M$ processes lodging in the memory while in the other two implementations, only one process is lodging in the memory. 
The memory consumption of a process consists of four parts: data ($D$), text ($E$), heap ($H$) and stack ($S$).
Data, text and heap sections are shared among threads and coroutines~\cite{silberschatz2018operating}. 
Therefore, in multi-process implementation, the total memory consumption ($C_1$) is the sum of $M$ data sections, $M$ text sections, $M$ heaps and $M$ stacks, or
\[C_{1}=M(D+E+H+S)\]
On the other hand, in the other two implementations, the total memory consumption ($C_2$) is 1 data section, 1 text section, 1 heap and $M$ stacks. 
The threads or coroutines also need to allocate $M$ times more memory in the heap compared to the processes, that is to say,
\[C_{2}=M(H+S)+D+E\]
Hence, $C_1-C_2=(M-1)(D+E)$, showing that the memory consumption of multi-process implementation is much more than that of multi-thread or multi-coroutine implementation. 
The stacks of different subroutines also differ. 
Windows and Linux allocate a few MB for a process or a thread, while Golang only allocates a few KB for a \textit{goroutine}. 
Under our time-series scenario where subroutines are simply ingesting data, the few KB are enough to finish the work. 
As a result, the memory consumption of a \textit{goroutine} is further less than a thread. 
In addition, the operating system needs Process Control Block (PCB) and Thread Control Block (TCB) to schedule processes and threads, while a program with coroutines also needs a similar data structure to schedule coroutines.
Therefore, the resource consumption of scheduling subroutines is nearly the same. 
In conclusion, the resource consumption of a process is greater than that of a thread; and that of a thread is greater than that of a coroutine. 

Second, consider the case that $M>N$, which is when the scheduling of subroutines starts to effect the concurrent system.
Most operating systems use time slice algorithm that originates from multi-batch systems to schedule processes and threads, i.e., the operating system allocates the active process or thread a certain time slice $t_1$.
Once the time slice is expired, the process of thread is forced to halt and a new one is activated~\cite{DBLP:conf/sosp/Dijkstra67}. 
The switch also needs time $t_2$ including saving and reloading contexts. 
Hence, the CPU utilization of running subroutines can be calculated as:
\[\eta=\frac{t_1}{t_1+t_2}\]
Therefore, the less time spent scheduling the subroutines, the more efficient a concurrent system is. 
Unfortunately, the switch of processes and threads is a type of system call, that is to say, the operating system must change from user mode to kernel mode, which consumes time. 
The scheduling of \textit{goroutines} does not involve the kernel mode and is completed under user mode, so it is faster to switch \textit{goroutines} than switch processes or threads. 
Bendersky conducted an experiment, which shows that the cost of Linux's context switch is about 2000ns, and that of \textit{goroutine} is about 170ns, which is near one twelfth of threads~\cite{bendersky2018}.

In our time-series scenario, \textit{goroutines} is further superior to processes and threads. 
It should be noted that multiple \textit{goroutines} can share one physical thread, and the creation and destruction of \textit{goroutines} cost far less than threads and processes, because it does not create or destroy physical threads or processes. 
Under our scenario, the main work of subroutines is to receive and transfer data, which is a kind of I/O operation. 
I/O operations are far more slower than CPU operations, so once a subroutine starts the I/O operation, it will halt itself to release resources to other subroutines, which makes $t_1$ reduce. 
Under normal cases, $t_1$ can be tens of milliseconds, but in our scenario $t_1$ can be several microseconds. 
To make the matter worse, we may start a lot of such subroutines in a short time to deal with the fast data flow. 
In multi-process or multi-thread implementations, the cost of creating and scheduling processes or threads is enormous given the amount of subroutines, and the CPU utilization can be reduced dramatically. 
Using the data from~\cite{bendersky2018}, when $t_1=10\text{ms}$, the CPU utilization of both Linux threads and \textit{goroutines} is almost 100\%. 
However, when $t_1=5\mu\text{s}$, the CPU utilization of Linux threads is 71.4\% and that of \textit{goroutines} is 96.7\%, which is significantly higher than multi-thread implementation. 
In conclusion, the time cost of creating, scheduling and destructing \textit{goroutines} are far less than that of processes and threads.

To sum up, the resource consumption and scheduling cost of \textit{goroutines} are far less than those of processes and threads. 
The two advantages make \textit{goroutines} able to reach a much higher DOP with less resource consumption and more CPU utilization.
Higher DOP means higher throughput of ingesting data.

\subsection{Lock-free queue}
\label{subsec:lockfree}
As we know in Section~\ref{sec:matrixGate}, both modules contains multiple coroutines. 
The coroutines are not in one-to-one correspondence, as the numbers of the coroutines in the two modules can be different. 
Thus, we need the data communication pipeline to connect the two modules. 
The coroutines in the \textit{data injection module} write data into the data communication pipeline, from which the coroutines in the \textit{data connection service module} read data. 
This means that the data communication pipeline is basically a queue, or specifically a concurrent FIFO queues.

\begin{figure}[htbp]
    \centering
    \includegraphics[width=.4\textwidth]{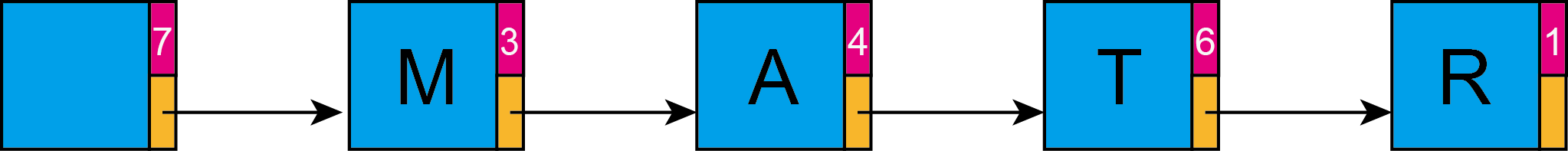}
    \caption{The structure of a sample lock-free queue.}
    \label{fig:queue}
\end{figure}

MatrixGate uses a lock-free queue based on the work of Michael and Scott~\cite{DBLP:conf/podc/MichaelS96} to implement the data communication pipeline. 
The queue is implemented by linked list, whose structure is shown in Figure~\ref{fig:queue}. 
The figure shows a queue containing four characters 'M', 'A', 'T' and 'R'. 
The blue region of a node represents its data, and the first node whose data is null is a dummy node. 
The yellow region represents the \texttt{next} pointer which points to the next node. 
We will introduce the pink region later.

The concurrency algorithm is based on \texttt{compare\_and\_swap} (CAS). 
CAS is an atomic operation that receives three parameters: shared address, expected value and new value. 
The operation compares the value of the shared address to the expected value. 
If they are equal, it assigns the new value to the shared address, before it returns if the operation has done. 
The naive CAS faces ABA problem. 
Before executing CAS, a program has to read the shared address to get the expected value A. 
If some other program modifies the value to B and then to A between the read and the CAS, the CAS will still be done.
To solve the problem, the algorithm adds a modification counter (the pink region in Figure~\ref{fig:queue}) to the shared address which records how many times the address is modified. 
Only when both the counter and the value are expected can a CAS be done. 
The algorithm uses this enhanced CAS to connect new nodes to the linked list (enqueue) and discard nodes from the list (dequeue). 
This algorithm outperforms other algorithms, cutting at least 1/3 of net execution time under high DOP.
More details about the lock-free queue can be seen in~\cite{DBLP:conf/podc/MichaelS96}.

\section{Experiments}
\label{sec:experiment}
We have conducted extensive experiments to show superiority of our MatrixGate over other data ingestion approaches. 
We first introduce our experiment design in Section~\ref{subsec:expdesign} and settings in Section~\ref{subsec:expsetting}. 
Then we evaluate our proposed solution and analyze experiment results in Section~\ref{subsec:load} to~\ref{subsec:query}.

\subsection{Experiment Design}
\label{subsec:expdesign}
Our proposed solution, MatrixGate, introduces a new ingestion strategy and utilizes multi-coroutine and lock-free queues, thus streamlining the data ingestion process and eliminating unnecessary intermediate steps, but there are still following questions need to be addressed:
\begin{enumerate}
\item How does MatrixGate perform compared to other data ingestion approaches? 
In other words, does MatrixGate ingest data faster?
\item Can MatrixGate complete common queries in time after data entry under time-series scenario?
\item How does MatrixGate scale as the cluster size increases? How much can the ingestion speed be improved?
\end{enumerate}

To answer these three questions, we design experiments in three dimensions. 
First, we compare MatrixGate with INSERT, COPY, \textit{gpfdist} on MatrixDB database platform by ingesting different datasets, to evaluate MatrixGate against previous methodologies.
Similarly, we also compare MatrixDB using MatrixGate with InfluxDB, QuestDB, TDengine and other state-of-the-art time-series databases by ingesting different datasets.
Second, we conduct experiments on the trade-off between ingesting and querying. 
There are three types of queries that are of most concern under time-series scenario, namely, the latest value, the detailed value within a certain period and aggregation. 
This experiment designs a query statement that queries the latest value of all devices throughout the time period, so that we test timeliness of MatrixGate.
Third, we conduct tests on different cluster sizes. 
In addition to MatrixGate, TDengine and ClickHouse, which are distributed time-series databases, are also included to the test so that we evaluate the horizontal scalability of MatrixGate.

\subsection{Experiment Settings}
\label{subsec:expsetting}
\subsubsection{Baselines}
\label{sss:baseline}
We compare MatrixGate with the data ingestion approaches supported by the databases listed below. 

\noindent{\normalsize\bfseries{MatrixDB 4.6.0}}~\cite{matrixdb}: MatrixDB is a hyper-converged distributed database supporting HTAP and IoT time-series applications, and supports data ingestion with INSERT, COPY, \textit{gpfdist}, and MatrixGate.

\noindent{\normalsize\bfseries{InfluxDB 1.8.10}}~\cite{Influxdb}: InfluxDB is a high-performance time-series database, ranked \#1 in DB-Engines' time-series database popularity ranking ~\cite{tsdbrank}, which supports using InfluxQL to write data directly, and writing data points to InfluxDB via \textit{InfluxDB line protocol}. 
Users can use import command to load data in local files.

\noindent{\normalsize\bfseries{TimescaleDB 2.7.2}}~\cite{timescaledb}: TimescaleDB is a time-series database that exists as an extension of PostgreSQL.
In addition to inserting data in the form of PostgreSQL's INSERT and COPY, it also implements a unique \textit{timescaledb-parallel-copy}~\cite{timescaledb/parallel/copy} ingestion approach, an SIPP optimization of COPY, to load time-series data more efficiently.

\noindent{\normalsize\bfseries{QuestDB 6.4.3}}~\cite{questdb}: QuestDB is a high-performance relational columnar storage time-series database that uses append-store columnar storage model where files are mapped to memory via a memory mapping table and each data write operation will append data at the end of mapped memory directly.
It is very efficient and supports ingesting data that meets InfluxDB Line Protocol or PostgreSQL wire protocol.
It provides data ingestion approaches for Web Console and HTTP REST API, and for local data, users can also use COPY approach to load data.

\noindent{\normalsize\bfseries{ClickHouse 22.7.2.15-1}}~\cite{clickhouse}: Clickhouse is a columnar storage distributed database for OLAP. 
Although it is not designed as a time-series database, it is also often used as a time-series database due to its high performance on data ingestion and query processing~\cite{struckov2019evaluation}. 
ClickHouse performs data insertion via the INSERT command, which not only allows single-point data insertion, but also supports ingestion from file or object stores.

\noindent{\normalsize\bfseries{TDengine 2.6.0.10}}~\cite{tdengine}: TDengine is a high-performance distributed relational time-series database that supports insertion of data points directly or from files via the INSERT command, as well as schema-free data writing through InfluxDB Line Protocol.

\noindent{\normalsize\bfseries{Druid 0.23.0}}~\cite{druid}: Apache Druid is an analytic time-series database that can load data from various data sources. 
In addition to inserting data points directly via INSERT and ingesting data from local files, it also supports ingesting streaming data from Kafka and Apache Kinesis, and ingesting batch data from HDFS and Amazon S3. 
Apache Druid provides a Web UI interface, through which users can configure and commit ingestion tasks.

In summary, Table~\ref{tab:baseline} shows our baselines.

\begin{table*}[!htb]
\centering
\caption{Baselines.}
\label{tab:baseline}
\begin{tabular}{ccccc}
\hline
Database & INSERT-like & COPY/LOAD-like & SIPP & MIPP  \\ \hline
MatrixDB & \checkmark & \checkmark & - & \textit{gpfdist} \\
InfluxDB & \checkmark & \checkmark & - & - \\
TimescaleDB & \checkmark & \checkmark & \textit{timescaledb-parallel-copy} & - \\
QuestDB & \checkmark & \checkmark & REST API & - \\
ClickHouse & \checkmark & \checkmark & insert from & - \\
TDengine & \checkmark & \checkmark & insert into & - \\
Apache Druid & \checkmark & \checkmark & Native batch & - \\ \hline
\end{tabular}
\end{table*}

\subsubsection{Datasets}
\label{sss:dataset}
A total of four datasets are used, including Stock~\cite{stock}, Weather~\cite{weather}, CPU-Only and IoT~\cite{tsbs}, which correspond to financial transactions, environmental monitoring, device monitoring and IoT scenarios respectively.
The statistics of all datasets are shown in Table~\ref{tab:statisticsOfDatasets}. 
All of the datasets are prestored as data files.

\begin{table*}[!htb]
\centering
\caption{Statistics of Datasets.}
\label{tab:statisticsOfDatasets}
\begin{tabular}{ccccc}
\hline
Datasets & Application Scenarios & Data Collection Approaches & No. of Rows & No. of Values \\\hline
Stock & Financial trading & 4259 stock daily prices from 1970 to 2021 & 24,510,593 & 194,571,142 \\\
Weather & Environmental Monitoring & 2000 locations each sampled 20000 times & 40,000,000 & 160,000,000 \\\
CPU-Only & Device Monitoring & 100,000 devices sampled every 10 seconds for 1 hour & 36,000,000 & 7,200,000,000 \\\
IoT & IoT & 1 million devices sampled every 10 seconds for 1 hour & 292,339,393 & 4,677,430,288 \\\hline
\end{tabular}
\end{table*}

\subsubsection{Performance Metrics}
\label{sss:metrics}
We evaluate different databases from two perspectives: ingestion speed and scalability. 
As for timely query experiment, we measure query latency to evaluate the database.

\noindent{\normalsize\bfseries{Ingestion speed}}: We define the ingestion speed $V$ as the throughput of data rows per unit time. 
Denote the total number of rows loaded as $N$ and the time spent on data ingestion as $T$. 
Given a database cluster with $K$ sub-instances, load speed $V$ can be calculated by formula ~\ref{eq:V}:
\begin{equation}
V=\frac{N}{T}=\frac{N}{\max(te_i-ts)}, i=1,…,K  \label{eq:V}
\end{equation}
where $ts$ denotes the moment when an ingestion task is started, and $te_i$ denotes the moment when the $i$-th database sub-instance completes the ingestion task. 
Here, completion of a ingestion task means that loaded data can be queried by the user through the database client.

\noindent{\normalsize\bfseries{Scalability}}: We define the scalability of $P$ as the ratio of increment of ingestion speed to increment of physical nodes of the database cluster, and use $P_{i,j}$ to denote the scaling efficiency when the number of physical nodes of the database cluster increases from $i$ to $j$. 
The calculation of $P_{i,j}$ is given in formula~\ref{eq:Pij}:
\begin{equation}
P_{i,j}=\frac{V_j}{V_i}/\frac{j}{i}=\frac{i V_j}{j V_i}, i < j  \label{eq:Pij}
\end{equation}
where $V_i$ denotes the ingestion speed when the number of database cluster nodes is $i$.

\noindent{\normalsize\bfseries{Query latency}}: We define query latency $D$ as the time difference between the moment $ds$, when the query is initiated locally from the database, and the moment $de$, when the database returns the query result.
The value of query latency is given in formula~\ref{eq:D}:
\begin{equation}
D=de-ds \label{eq:D}
\end{equation}

\subsubsection{Environment}
All experiments are conducted with r5.8xlarge instances of AWS EC2~\cite{ec2/r5}, and the specific configuration information is shown in table~\ref{tab:machineDetail}.

\begin{table}[!htb]
\centering
\caption{Environment Configuration.}
\label{tab:machineDetail}
\begin{tabular}{ll}
\hline
Configuration & Detail \\ \hline
OS & CentOS-7.20009 \\
OS AMI ID & 0fef8c1409596e79f \\
CPU & Intel Xeon Platinum 8000 \\
vCPU & 32 \\
Clock Frequency & 3.1GHz \\
Network Bandwidth & 10Gigabit \\
Memory & 256GB \\
Storage volumes & 1000GiB, 10000IOPS, 1000MiB/s \\ \hline
\end{tabular}
\end{table}

The same configuration is used for both single-node and multi-node (up to 5 nodes) tests, and on-demand instances are started in \textit{ap-southeast-1a} availability zone for data ingestion tests.

\subsection{Performance}
\label{subsec:load}
For the first question in Section~\ref{subsec:expdesign}, in order to verify the performance improvement brought by MatrixGate's combination of SIPP and MIPP, we first compare MatrixGate with previous data ingestion methodologies, including INSERT, COPY, and \textit{gpfdist}, on a single-node MatrixDB by ingesting different datasets.
We ensure that the table structure, storage engine (heap tables) and database parameter configuration are the same, except for data ingestion approaches.

\begin{figure}[!htb]
    \centering
    \includegraphics[width=0.4\textwidth]{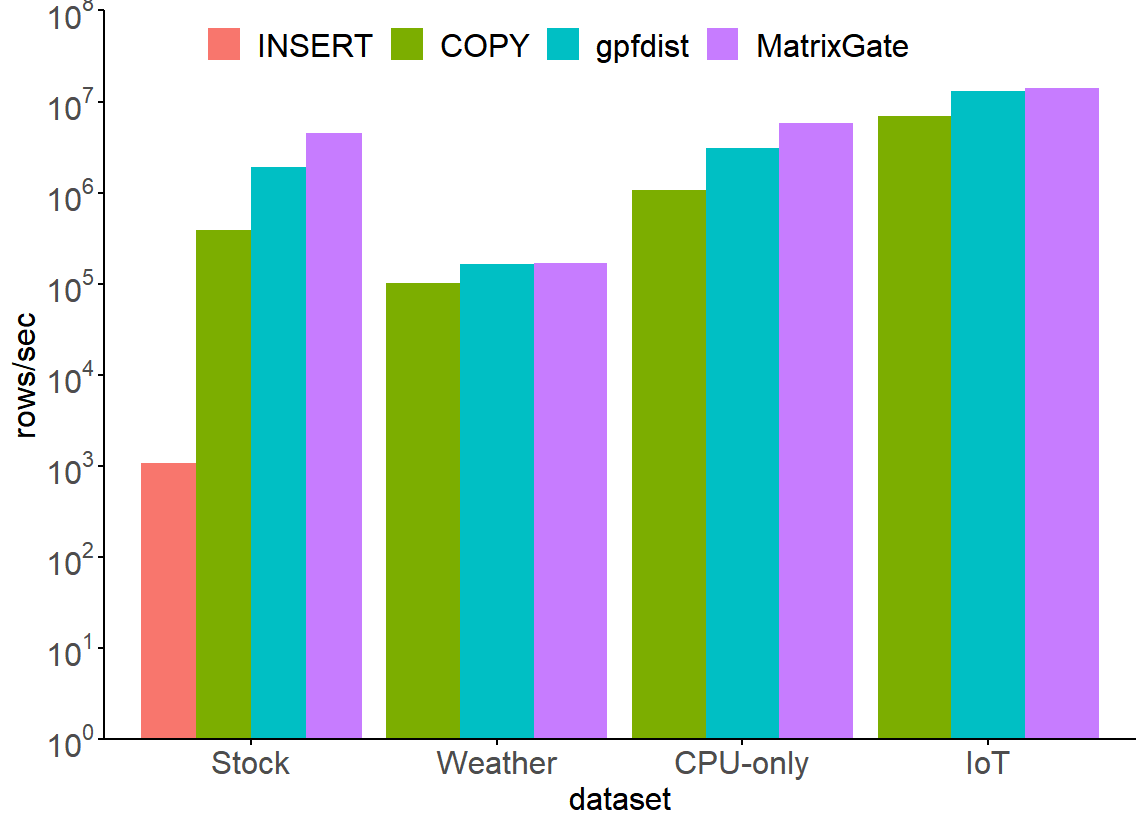}
    \caption{Improvement of MatrixGate compared to traditional methodologies.}
    \label{fig:insert to mxgate}
\end{figure}

\subsubsection{Ingestion approaches}
As shown in Figure ~\ref{fig:insert to mxgate}, we test four data ingestion approaches, INSERT, COPY, gpfdist and MatrixGate, on four datasets. 
Since the direct insertion of data points via INSERT is too slow and takes too long, INSERT is only experimented on the Stock dataset. 
According to the experimental results, the ingestion speed has improved about 365 times from INSERT with direct insertion of data points to COPY with batch processing, which is a significant improvement due to transaction simplification and parsing optimization. 
The improvement from COPY to \textit{gpfdist} ranges roughly from 2.5 times to 4.8 times. 
Note that although the test environment is single-node and it is actually deployed with 6 instances, enabling the optimization of \textit{gpfdist} and resulting in several times of performance improvement. 
Finally, the ingestion speed of MatrixGate is 1.4 to 2.4 times faster than \textit{gpfdist}. 
MatrixGate incorporates SIPP optimization compared to \textit{gpfdist}, but the bottleneck exists due to restricted number of cores on the server, preventing MatrixGate from reaching a higher performance improvement. 
If the the ratio of CPU core to sub-instance could be further improved, MatrixGate should have a higher performance improvement than \textit{gpfdist}.

\subsubsection{Ingestion speed vs DOP}
\begin{figure}[!htb]
    \centering
    \includegraphics[width=0.4\textwidth]{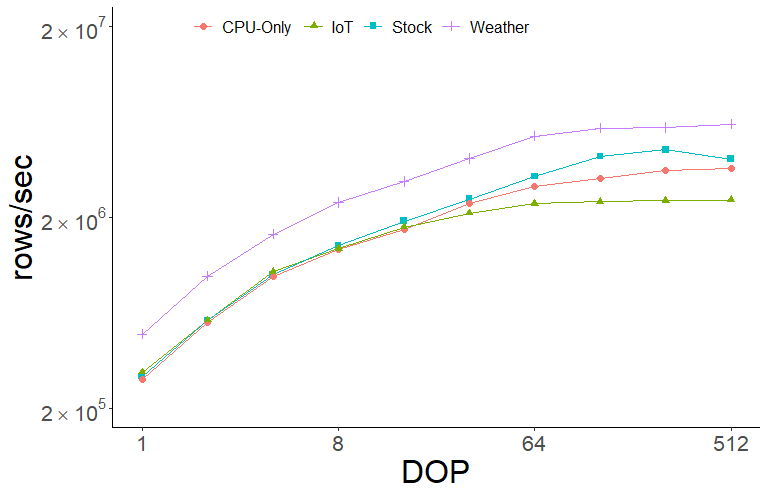}
    \caption{Ingestion speed of MatrixGate for different DOPs.}
    \label{fig:increase}
\end{figure}

While conducting experiments on comparison of data ingestion approaches, we also collect the trend of MatrixGate's ingestion speed as DOP increases, as shown in Figure ~\ref{fig:increase}.
Across the four datasets, there is a clear pattern in the ingestion performance as DOP increases. 
Initially, there is a steep increase in ingestion speed as DOP grows from 1 to 64, with the speed almost doubling whenever DOP is doubled, demonstrating an efficient parallelization within this range.
However, post-DOP 64, the increase in ingestion speed begins to plateau, converging towards a stable value as DOP is extended to 512. 
This behavior is the indicative of the server's physical limitations, where the number of CPU cores and available memory space restrict the efficacy of adding more parallel processes. 
Beyond a certain point, the overhead associated with managing increased parallelism, such as context switching and scheduling, begins to outweigh the benefits, potentially leading to a stagnation or even decline in ingestion speed. 
The dataset of particular interest is Stock, which is the smallest dataset with approximately 24 million rows, and exhibits a noticeable drop in ingestion speed at a DOP of 512. 
This drop suggests that the level of parallelism has surpassed the optimal point for the size of the dataset. 
With an abundance of coroutines for such a relatively small dataset, the additional overhead caused by extra coroutines and slots does not contribute to performance gains but instead leads to inefficiency. 
The trend lines for all datasets start to converge as DOP increases, yet it is especially pronounced for the Stock dataset, underscoring that while parallelism can significantly improve performance, there is a threshold beyond which further parallelization is counterproductive. 
This limit is influenced by the dataset size and the hardware constraints of the server.

\subsubsection{Single-node ingestion speed}
We continue to complete single-node deployments of MatrixDB, InfluxDB, TimescaleDB, QuestDB, ClickHouse, TDengine, and Apache Druid on identically configured servers, and conduct data ingestion experiments on four datasets to compare ingestion speeds. 
Specifically, InfluxDB is a schema-less database that automatically creates time-series data fields based on the loaded data, using the TSM Tree storage engine; TimescaleDB uses a hypertable for time-series data; MatrixDB and QuestDB use columnar append tables for time-series data; ClickHouse uses MergeTree, a columnar storage engine for fast data writing; TDengine uses an LSM-like TSDB storage engine; Apache Druid automatically recognizes data patterns and uses a columnar storage engine. 
Among them, MatrixDB and ClickHouse support a variety of storage engines. 
Since we focus on data ingestion performance, we only choose the storage engine suitable for fast time-series data writing. 
Under practical application scenarios, it is crucial to choose the appropriate storage engine for superior performance.

\begin{figure}[!htb]
    \centering
    \includegraphics[width=0.4\textwidth]{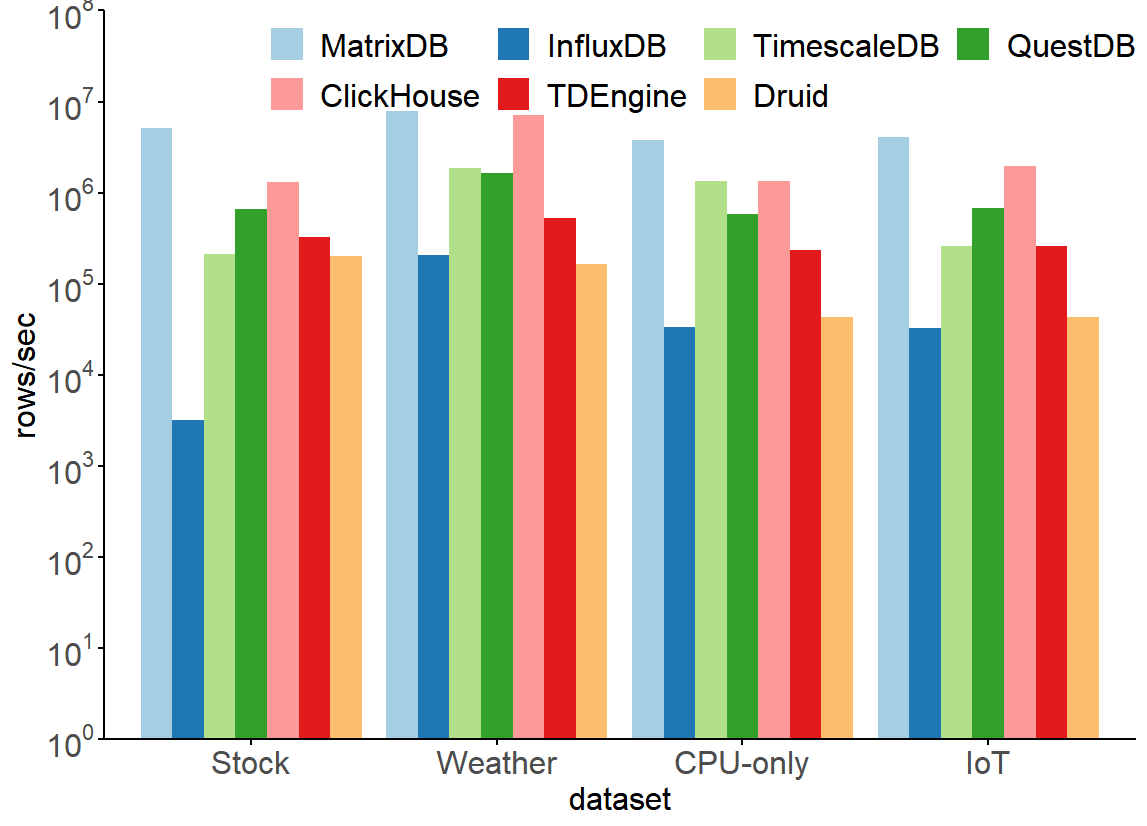}
    \caption{Ingestion speed of different time-series databases.}
    \label{fig:throughput}
\end{figure}

The experimental results are shown in Figure~\ref{fig:throughput}, with different colors representing different systems. 
In the bar chart, different approaches are identified by databases, such as MatrixDB corresponding to MatrixGate, TimescaleDB corresponding to \textit{timescaledb-parallel-copy}, with other databases corresponding to time-series data ingestion approaches described in Section~\ref{sss:baseline}. 
MatrixDB, identified by the color light blue, outperforms all other databases across the board. 
It achieves peak ingestion rates of 7.89 million rows per second on the Weather dataset and 5.13 million rows per second on the Stock dataset. 
For the CPU-Only and IoT datasets, which are characterized by more complex data patterns and larger volumes, MatrixDB sustains high ingestion rates of 3.66 million rows per second and 3.75 million rows per second, respectively. 
Closest to MatrixDB's performance is ClickHouse, denoted by the light red bars. 
ClickHouse nearly matches MatrixDB on the Weather dataset with a throughput of 7.19 million rows per second but falls behind on the other datasets. 
Its performance on the Stock, CPU-Only, and IoT datasets is recorded at 1.31 million, 1.34 million, and 1.96 million rows per second, respectively—indicating approximately one-third to one-half the ingestion rate of MatrixDB. 
TimescaleDB and QuestDB, shown in light green and dark green respectively, also demonstrate the capability to ingest millions of rows per second in certain conditions. 
However, their performance varies with the dataset. 
TimescaleDB excels in handling the CPU-Only and Weather datasets, while QuestDB shows stronger performance with the Stock and IoT datasets, emphasizing their specialized efficiencies in handling specific types of data. 
TDengine, represented by the dark red bars, maintains a consistent ingestion rate in the range of hundreds of thousands of rows per second across all four datasets. 
In contrast, Apache Druid and InfluxDB, indicated by yellow and dark blue bars respectively, exhibit the slowest ingestion rates among the databases evaluated. 
Both manage to process only tens of thousands of rows per second, which is a mere fraction—approximately one percent—of MatrixDB's throughput. 
This chart not only demonstrates the superiority of MatrixDB's MatrixGate in data ingestion but also highlights the diverse performance profiles of each database system, suggesting their potential best-fit scenarios based on the nature of the data being ingested.

The reason for this comparative result is that, as described in previous sections, a good parallel strategy can lead to great enhancement of performance. 
MatrixGate, ClickHouse's INSERT and TimescaleDB's \textit{timescaledb-parallel-copy}, all implement SIPP, while MatrixDB and ClickHouse support columnar storage engine, which can match data ingestion tools better to achieve higher ingestion performance. 
Besides, MatrixGate features direct data transfer in addition to parallel optimization, thus achieving optimal ingestion performance. 
This reveals that MatrixGate is better able to meet the challenges of efficient time-series data ingestion.

\subsection{Timeliness}
\label{subsec:query}
For the second question in Section~\ref{subsec:expdesign}, we did a query latency experiment.
Data ingestion and query are closely related to the design of the storage engine. 
MatrixGate's efficient data ingestion is achieved with the cooperation of MatrixDB's columnar storage engine, which needs to handle the balance between efficient ingestion and efficient query. 
Therefore, we conduct typical query experiments on InfluxDB and MatrixGate (which uses MatrixDB's columnar append-table storage engine), to observe how these two storage engines handle this problem. 
The following SQL statements are executed on the IoT dataset\footnote{Note that the syntax of SQL statements in two databases is different, and two quotes of SQL statements in the databases can prove to be literally equivalent.}.

\begin{lstlisting}[ language=SQL,
                    deletekeywords={IDENTITY},
                    deletekeywords={[2]INT},
                    morekeywords={clustered},
                    framesep=8pt,
                    xleftmargin=40pt,
                    framexleftmargin=40pt,
                    frame=tb,
                    framerule=0pt ]
-- MatrixDB
SELECT 
  name, last(driver, ts) as driver, 
  last(longitude, ts) as longitude,
  last(latitude, ts) as latitude 
FROM readings r 
GROUP BY name;
-- InfluxDB
SELECT 
  "name", "driver", "longitude", "latitude" 
FROM "readings" 
GROUP BY "name", "driver" 
ORDER BY "time" 
LIMIT  1;
\end{lstlisting}

\begin{figure}[!htb]
    \centering
    \includegraphics[width=0.4\textwidth]{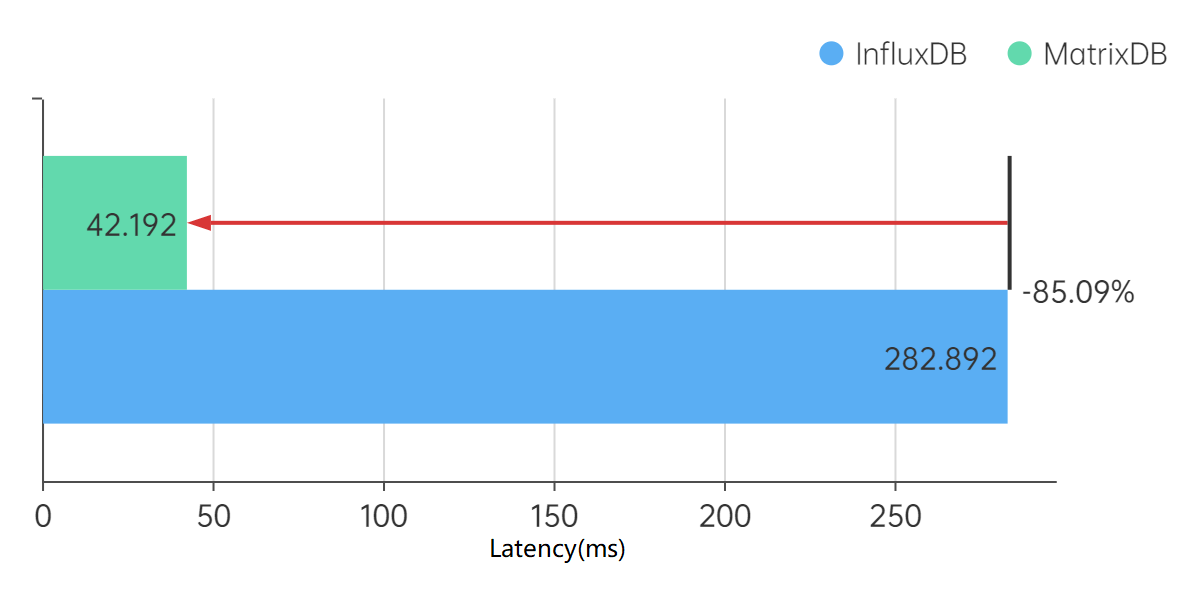}
    \caption{Query latency.}
    \label{fig:query}
\end{figure}

The test result is shown in Figure~\ref{fig:query}. 
MatrixDB completes such a representative query very efficiently, taking only $14.91\%$ of InfluxDB's time. 
The reason why MatrixDB can complete this typical query so quick is that \verb|last|, a time-series function, is sufficiently optimized specifically for time-series scenarios in MatrixDB's implementation.
This test illustrates that MatrixGate's efficient data ingestion has no significant negative impact on the query tasks after data entry, and can be used in practice.

\subsection{Scalability}
\label{subsec:scalability}
For the third question in Section~\ref{subsec:expdesign}, as a system designed for distributed architecture, distributed deployment of the database cluster can truly leverage the capability of MatrixGate, so we conduct horizontal scalability experiments. 
In this experiment, we choose the most challenging IoT dataset in the previous experiments, and test data ingestion approaches of MatrixDB, ClickHouse and TDengine, which perform well in Section \ref{subsec:load} and support distributed data ingestion, with MatrixDB using columnar append tables, ClickHouse using MergeTree storage engine, and TDengine using LSM-like TSDB storage engine.
Constrained by the server resources, we only conduct experiments on 1, 3 and 5 nodes.

\begin{figure}[!htb]
    \centering
    \includegraphics[width=0.4\textwidth]{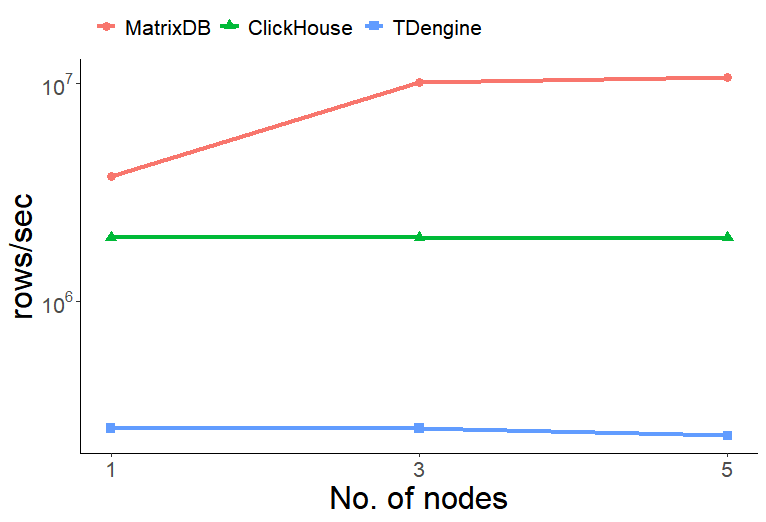}
    \caption{Ingestion speed for different nodes.}
    \label{fig:distributed database}
\end{figure}

As shown in Figure ~\ref{fig:distributed database}, the ingestion speed of MatrixGate increases from 3.75 million rows/sec to 10.05 million rows/sec as the number of nodes in MatrixDB increases from 1 to 3, with a performance improvement of about 2.68 times. 
The scalability can be calculated as $2.68/3 \approx 89 \%$, which is near-linear.

Note that a distributed system with good, or even linear scalability is unable to scale infinitely in practice. 
In distributed systems with a larger number of nodes, other factors, such as network bandwidth, communication between nodes and hardware configuration, may become the bottleneck and restrict the real scalability of systems. 
Figure~\ref{fig:distributed database} also shows that the data ingestion speed of MatrixGate stagnates with more than five nodes. 
This is because our tested time-series data is stored as files, and MatrixGate has to read files via \textit{cat} command and pipeline, which makes file system the new bottleneck. 
In the case of 3 nodes, the throughput of MatrixGate reaches about 10G/sec, near the bandwidth of the server, and thus MatrixGate fails to increase its throughput in the case of 5 nodes. 
There are 2 reasons why we use files to conduct experiments. 
On the one hand, randomly generated data cannot characterize time-series data, and thus we need to use historical time-series data; on the other hand, ingestion via files can not only avoid the overhead of generating data, but also have our experiments repeatable. 

In contrast, the graphs of ClickHouse and TDengine are essentially horizontal lines, and their ingestion performances do not improve with the increase in the number of cluster nodes. 
This is because ClickHouse and TDengine do not implement MIPP, and data ingestion through a single access point is essentially equivalent to single-node data ingestion, presenting horizontal lines. 
The database clusters of ClickHouse and TDengine use a peer-to-peer architecture.
It is also possible to manually slice and dice the data and then load data slices through multiple access points of the database cluster at the same time, but this approach is more complex to operate and not as easy to use as MatrixGate.

\begin{figure}[!htb]
    \centering
    \includegraphics[width=0.4\textwidth]{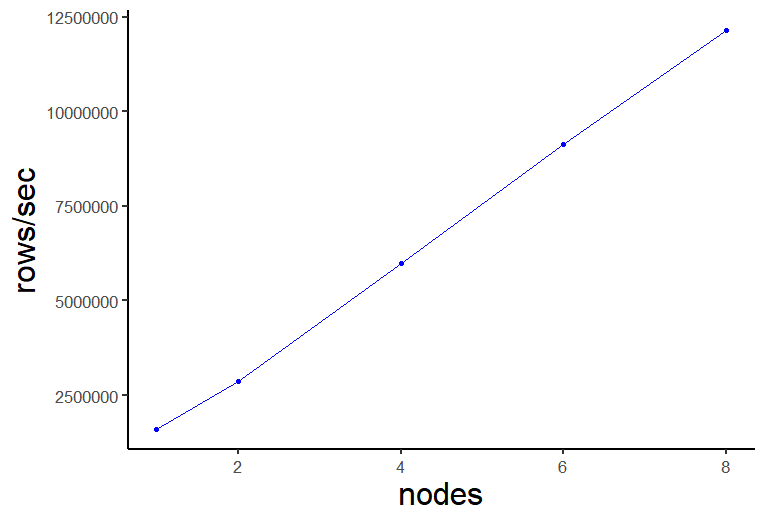}
    \caption{MatrixGate's ingestion speed under heavy load scenario.}
    \label{fig:scalability}
\end{figure}

We also conduct experiments to investigate the performance of a server executing queries and ingesting data simultaneously, which can introduce constraints on the computational resources available for data ingestion and limit the maximum data ingestion speed on each node. 
In contrast to the other experiments conducted on the r5.8xlarge instance, this particular experiment is performed on eight AWS EC2 r6i.2xlarge instances. 
Building on discussions above, this experiment provides additional evidence that MatrixGate is a highly scalable solution. 
Figure~\ref{fig:scalability} shows the result of the experiment and clearly demonstrates the impressive scalability of MatrixGate. 
Specifically, when the number of nodes increases from 1 to 8, the ingestion speed increases by approximately 7.59 times, resulting in its scalability of $94.9\%$, which is astonishing.

In summary, MatrixGate has good horizontal scalability.
Increasing the number of cluster nodes can further improve the ingestion speed of MatrixGate, so it can better cope with the challenges of ingesting time-series data caused by the rapid growth of time-series data.

\section{Related Work}
\label{sec:relatedWork}
In this section, we will introduce the related work on SIPP and MIPP.

\noindent{\normalsize\bfseries{SIPP}}: 
Examples of SIPP includes the modified INSERT in ClickHouse and \textit{timescaledb-parallel-copy} of TimeScaleDB. 
The COPY and INSERT commands are implemented in SISP by traditional database systems. 
By modifying these commands, TimeScaleDB and ClickHouse can load data in parallel through simply calling COPY and INSERT commands. 
\textit{Timescaledb-parallel-copy} takes the naive SIPP, dividing the data into chunks and loading the chunks in parallel~\cite{timescaledb/parallel/copy}. 
ClickHouse develops a high performance storage engine, MergeTree uses Log-Structured Merge-trees (LSM trees) to organize its data. 
While executing a modified INSERT command, the data is loaded into memory in parallel first, and then merged into LSM trees asynchronously and periodically~\cite{clickhouse/insert}. 
Both of the two approaches are based on files, so data cannot be written into database in a short time.

\noindent{\normalsize\bfseries{MIPP}}: 
A basic and naive approach of MIPP is COPY ON SEGMENT. 
The data is split into chunks either manually or semi-automatically, and then the chunks is distributed to the file systems of the sub-instances. 
The database cannot load data until these preparations are done~\cite{copy/on/segment}. 
InfluxDB clusters and ClickHouse support this approach. 
The preparations of this approach is neither simple nor fast, so it is hardly used. 
Another effective way is to utilize external tables. 
An external table allows user to query data outside the database, and the table can be located in the local file system, a cloud storage or even another database system~\cite{bryla2006ocp}. 
The mechanism eliminates the preparation of COPY ON SEGMENT and when the database is to load data, it simply starts an INSERT command like a normal query, and MIPP loader instances will load the target data in parallel. 
For example, \textit{gpfdist} of Greenplum utilizes the external table. 
During the execution of the INSERT command, the segments of all sub-instances will access the target data and load data in uniform~\cite{gpfdist}. 
\textit{Gpossext} of AnalyticDB also utilizes the external table~\cite{gpossext}. 
Other examples of utilizing external tables are given in~\cite{greenplumExternaltable,oracleExternaltable}.

\section{Conclusion}
\label{sec:conclusion}
We introduce MatrixGate, a high-performance time-series data ingestion approach based on multi-coroutine and lock-free queues. 
For both single-node and distributed deployments, MatrixGate can make high use of parallel capability of multi-core CPUs to increase loading speed. 
MatrixGate creatively implements direct data transfer, which allows the data to be processed faster, more efficiently and at a lower cost. 
It also solves the single-point performance bottleneck problem and supports parallel ingestion on multiple instances. 
In addition, MatrixGate provides real-time data entry with a hundred millisecond delay through micro-batch load and commit, which is not possible with previous time-series data ingestion approaches.


\bibliographystyle{ACM-Reference-Format}
\bibliography{main}

\end{document}